\documentclass[longauth]{aa} % for the long lists of affiliations 
\usepackage{graphicx,epsfig,subcaption,caption}
\usepackage{footnote,natbib}
\usepackage{txfonts}
\usepackage{longtable}
\usepackage{xcolor}	% color text
\begin{document} 

   \title{The supernova of the MAGIC GRB\,190114C\thanks{Based on data obtained under programs 199.D-0143(R) and 1103.D-0328(F) (PIs S. Smartt, C. Inserra), 18B-W18BN007 (PI K. M. L\'{o}pez), A38DDT3 (PI A. Melandri), 0102.D-0540(A) (PI E. Pian), 2019A-I0036-0 (PI A. Rossi), and 15684, 15708, \& 15979 (PI A. Levan) at the NTT, WHT, TNG, VLT, LBT, and {\it HST} telescopes, respectively.}}

\author{A. Melandri\inst{1}, L. Izzo\inst{2}, E. Pian\inst{3,4}, D. B. Malesani\inst{2}, M. Della Valle\inst{5,6,7}, A. Rossi\inst{3}, P. D'Avanzo\inst{1}, D. Guetta\inst{8,32}, P. A. Mazzali\inst{4,9}, S. Benetti\inst{10}, N. Masetti\inst{3,11}, E. Palazzi\inst{3}, S. Savaglio\inst{12}, L. Amati\inst{3}, L. A. Antonelli\inst{13}, C. Ashall\inst{14},  M. G. Bernardini\inst{1}, S. Campana\inst{1}, R. Carini\inst{13}, S. Covino\inst{1}, V. D'Elia\inst{13,15}, A.  de Ugarte Postigo\inst{16}, M. De Pasquale\inst{17}, A. V. Filippenko\inst{18,19}, A. S. Fruchter\inst{20}, J. P. U. Fynbo\inst{21}, A. Giunta\inst{15}, D. H. Hartmann\inst{22}, P. Jakobsson\inst{23}, J. Japelj\inst{24}, P. G. Jonker\inst{25,26}, D. A. Kann\inst{16}, G. P. Lamb\inst{27}, A. J. Levan\inst{25}, A. Martin-Carrillo\inst{28}, P. M{\o}ller\inst{29},  S. Piranomonte\inst{13}, G. Pugliese\inst{24}, R. Salvaterra\inst{30}, S. Schulze\inst{31}, R. L. C. Starling\inst{27}, L. Stella\inst{13}, G. Tagliaferri\inst{1}, N. Tanvir\inst{27}, D. Watson\inst{21}
          }
          
% Authors Institutes
%1
\institute{INAF - Osservatorio Astronomico di Brera, Via E. Bianchi 46, I-23807, Merate (LC), Italy\\
\email{andrea.melandri@inaf.it}
%2
\and DARK, Niels Bohr Institute, University of Copenhagen, Jagtvej 128, 2200 Copenhagen, Denmark
%DTU Space, National Space Institute, Technical University of Denmark, Elektrovej 327, 2800 Kongens Lyngby, Denmark
%3
\and INAF - Osservatorio di Astrofisica e Scienza dello Spazio, Via P. Gobetti 101, I-40129 Bologna, Italy
%4
\and Astrophysics Research Institute, Liverpool JMU, IC2, Liverpool Science Park, 146 Brownlow Hill, Liverpool L3 5RF, UK
%5
\and INAF - Osservatorio Astronomico di Capodimonte, Salita Moiariello 16, I-80131 Napoli, Italy
%6
\and INFN Napoli, Strada Comunale Cinthia, I-80126 Napoli NA
%7
\and ICRANet, Piazza della Repubblica 10, I-65122 Pescara, Italy
%8
\and Physics Department, University of Ariel, Ariel, West Bank, Israel
%ORT-Braude College, Snunit St 51, Karmiel, 2161002, Israel
%9
\and Max-Planck-Institut f\"ur Astrophysik, Karl-Schwarzschild Str. 1, D-85748 Garching, Germany
%10
\and INAF - Osservatorio Astronomico di Padova, Vicolo dell’Osservatorio 5, I-35122, Padova, Italy
%11
\and Departamento de Ciencias F\'isicas, Universidad Andr\'es Bello, Fern\'andez Concha 700, Las Condes, Santiago, Chile
%12
\and Physics Department, University of Calabria, I-87036, Arcavacata di Rende (Cs), Italy
%13
\and INAF - Osservatorio Astronomico di Roma, Via Frascati 33, I-00040 Monte Porzio Catone (RM), Italy
%14
\and Institute for Astronomy, University of Hawaii at Manoa, 2680 Woodlawn Drive Honolulu, HI 96822, USA
%15
\and Space Science Data Center (SSDC) - Agenzia Spaziale Italiana (ASI), Via del Politecnico, I-00133 Roma, Italy
%16
\and Instituto de Astrof\'{i}sica de Andaluc\'{i}a (IAA-CSIC), Glorieta  de la Astronom\'{i}a s/n, 18008 Granada, Spain
%17
\and Department of Astronomy and Space Sciences, Istanbul University, Beyazıt 34119, Istanbul, Turkey
%18
\and Department of Astronomy, University of California, Berkeley, CA, 94720-3411, USA
%19
\and Miller Institute for Basic Research in Science, University of California, Berkeley, CA 94720, USA
%20
\and Space Telescope Science Institute, 3700 San Martin Drive, Baltimore, MD 21218, USA
%21
\and The Cosmic Dawn Centre (DAWN), Niels Bohr Institute, University of Copenhagen, Lyngbyvej 2, 2100, Copenhagen, Denmark
%22
\and Department of Physics and Astronomy, Clemson University, Clemson, SC29634-0978, USA
%23
\and Centre for Astrophysics and Cosmology, Science Institute, Univer-sity of Iceland, Dunhagi 5, 107, Reykjavik, Iceland
%24
\and Anton Pannekoek Institute for Astronomy, University of Amsterdam, Science Park 904, 1098 XH Amsterdam, The Netherlands
%25
\and Department of Astrophysics, Radboud University, 6525 AJ Nijmegen, The Netherlands
%26
\and SRON,  Netherlands  Institute  for  Space  Research,  Sorbonnelaan 2, 3584 CA, Utrecht, The Netherlands
%27
\and School of Physics and Astronomy, University of Leicester, University Road, LE1 7RH, UK
%28
\and School of Physics, University College Dublin, Dublin 4, Ireland
%29
\and ESO Headquarters, Karl-Schwarzschildstrasse 2, D-85748, Garching, Germany
%30
\and INAF -- Istituto di Astrofisica Spaziale e Fisica Cosmica, Via A. Corti 12, I-20133, Milano, Italy
%31
\and Department  of  Particle  Physics  and  Astrophysics,  Weizmann Institute of Science, Rehovot 7610001, Israel
%32
\and Department of physics, ORT-Braude College, Snunit St 51, Karmiel, 2161002, Israel
             }

   \date{}

\titlerunning{The supernova of the MAGIC GRB\,190114C}
\authorrunning{Melandri et al.}

  \abstract
{We observed GRB\,190114C (redshift $z = 0.4245$), the first GRB ever detected at TeV energies,  at optical and near-infrared wavelengths with several ground-based telescopes and the {\it Hubble Space Telescope}, with the primary goal of studying its underlying supernova, SN\,2019jrj. The monitoring spanned the time interval between 1.3 and 370 days after the burst, in the observer frame. We find that the afterglow emission can be modelled with a forward shock propagating in a uniform medium modified by time-variable extinction along the line of sight. A jet break could be present after 7 rest-frame days, and accordingly the maximum luminosity of the underlying SN ranges between that of stripped-envelope core-collapse supernovae (SNe) of intermediate luminosity, and that of the luminous GRB-associated SN\,2013dx. The observed spectral absorption lines of SN\,2019jrj are not as broad as in classical GRB-SNe, and are rather more similar to those of less-luminous core-collapse SNe. Taking the broad-lined stripped-envelope core-collapse SN\,2004aw as an analogue, we tentatively derive the basic physical properties of SN\,2019jrj. We discuss the possibility that a fraction of the TeV emission of this source might have had a hadronic origin and estimate the expected high-energy neutrino detection level with IceCube. }
  

\maketitle

%
%-------------------------------------------------------------------

\section{Introduction}

GRB\,190114C was first identified as a long-duration gamma-ray burst (GRB) by the Burst Alert Telescope \citep[BAT;][]{bat} onboard the Neil Gehrels {\it Swift} Observatory \citep[{\it Swift;}][]{swift} and the Gamma-ray Burst Monitor \citep[GBM;][]{meegan09} of the {\it Fermi} satellite \citep{Swift,Fermi}. The trigger time was $T_{\rm 0}$ = 20:57:03 (UT dates are used throughout this paper). The time interval including 90\% of the flux ($T_{90}$) is $\sim 116$\,s as measured by {\it Fermi}/GBM (50--300\,keV energy band) and $\sim 362$\,s as measured by {\it Swift}/BAT (15--350\,keV energy band). The fluence as measured by GBM in the 10–1000 keV energy band is $(4.433 \pm 0.005) \times 10^{-4}$\,erg\,cm$^{-2}$ \citep{ajello20}.

GRB\,190114C is the first reported GRB that was also detected in the TeV band by The Major Atmospheric Gamma Imaging Cherenkov (MAGIC) telescopes. High-energy gamma-rays (0.2--1\,TeV) were reported by the MAGIC Collaboration with high significance from the beginning of the observations and lasted for at least 20\,min \citep{Magicnature1}; the emission was detected during both the prompt and afterglow phases. The emission component of the early afterglow has a comparable power to that of the standard synchrotron component and has been interpreted as being due to the inverse-Compton mechanism \citep{Magicnature2}. However, since no firm conclusions about the production mechanisms of the GeV--TeV emission have been reached so far \citep{ravasio19,fraija19b,wang19,derishev19,slf20,chand20,rueda20}, and since a pure leptonic scenario does not match the observed emission for GRB\,190114C, we cannot exclude the hypothesis that part of this GeV--TeV emission may be caused by the presence of a hadronic component (Gagliardini et al. 2021, in prep.), as we discuss in Section 4.

The event was also detected by the Large Area Telescope \citep[LAT;][]{atwood09} of the {\it Fermi} satellite \citep{FermiLAT}. By including the prompt and extended emission, the estimated fluence is $\sim 2.5 \times 10^{-5}$\,erg\,cm$^{-2}$ (100\,MeV--100\,GeV energy band), second only to GRB\,130427A \citep{ajello20}.

The afterglow emission from this GRB was detected at various wavebands from 0.65\,GHz to 23\,GeV \citep{Magicnature1,Magicnature2,laskar19,misra21,jm20,ajello20}. This allowed for the measurement of its redshift, $z \approx 0.4245$ \citep{redshift0,redshift}, and a detailed study of its host-galaxy properties \citep{deUgarte}. The uniqueness of GRB\,190114C motivated our search for the possible associated supernova (SN), despite the adverse circumstances of relatively large distance and high local background (the bright host galaxy and the line-of-sight extinction). In this paper, we present the results of our observational campaign in which we detected and identified a SN component underlying the GRB (\citealp[SN\,2019jrj;][]{mela19a,mela21}) and characterized its behavior.

Throughout the paper, distances are computed assuming a $\Lambda$CDM Universe with H$_{\rm 0} = 73$\,km\,s$^{-1}$\,Mpc$^{-1}$, $\Omega_{\rm m}$ = 0.27, and $\Omega_{\rm \Lambda}$ = 0.73 \citep{spergel07,riess16}. Magnitudes are in the AB system \citep{okegunn83} and uncertainties are at a 1$\sigma$ confidence level.

%--------------------------------------------------------------------
\section{Observations and data reduction}

   \begin{figure*}[!ht]
   \centering
    \includegraphics[width=18.4cm,height=6cm]{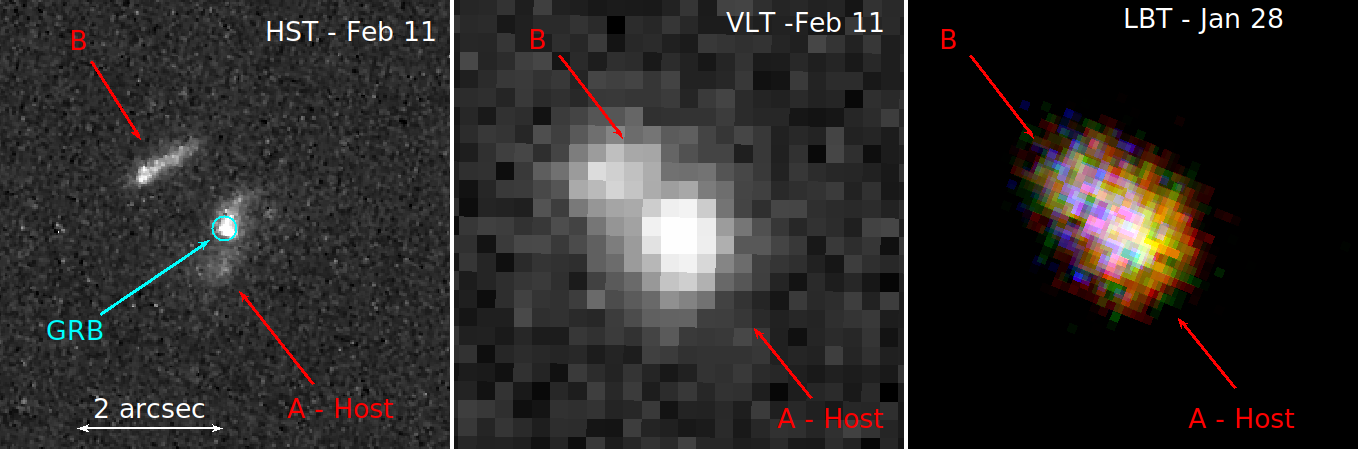}
   \caption{GRB\,190114C host galaxy (A) and its interacting galaxy (B), as observed with {\it HST} ($F606W$ filter), VLT ($R$ filter), and LBT ($gri$ image)  at similar epochs. North is up and east is to the left.}
    \label{figField}%
    \end{figure*}

   \begin{figure*}[!ht]
   \centering
     \includegraphics[width=8cm,height=8cm]{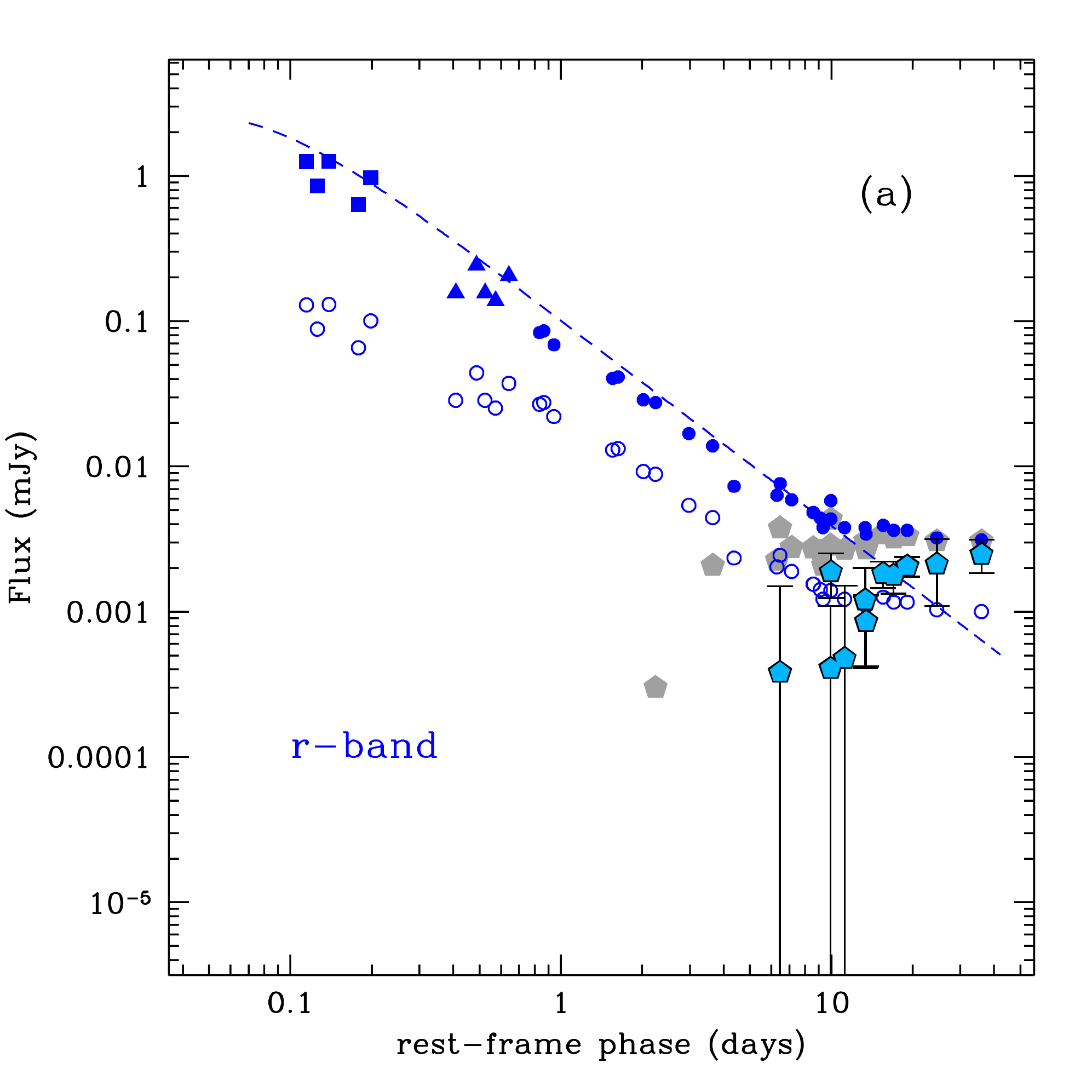}
     \includegraphics[width=8cm,height=8cm]{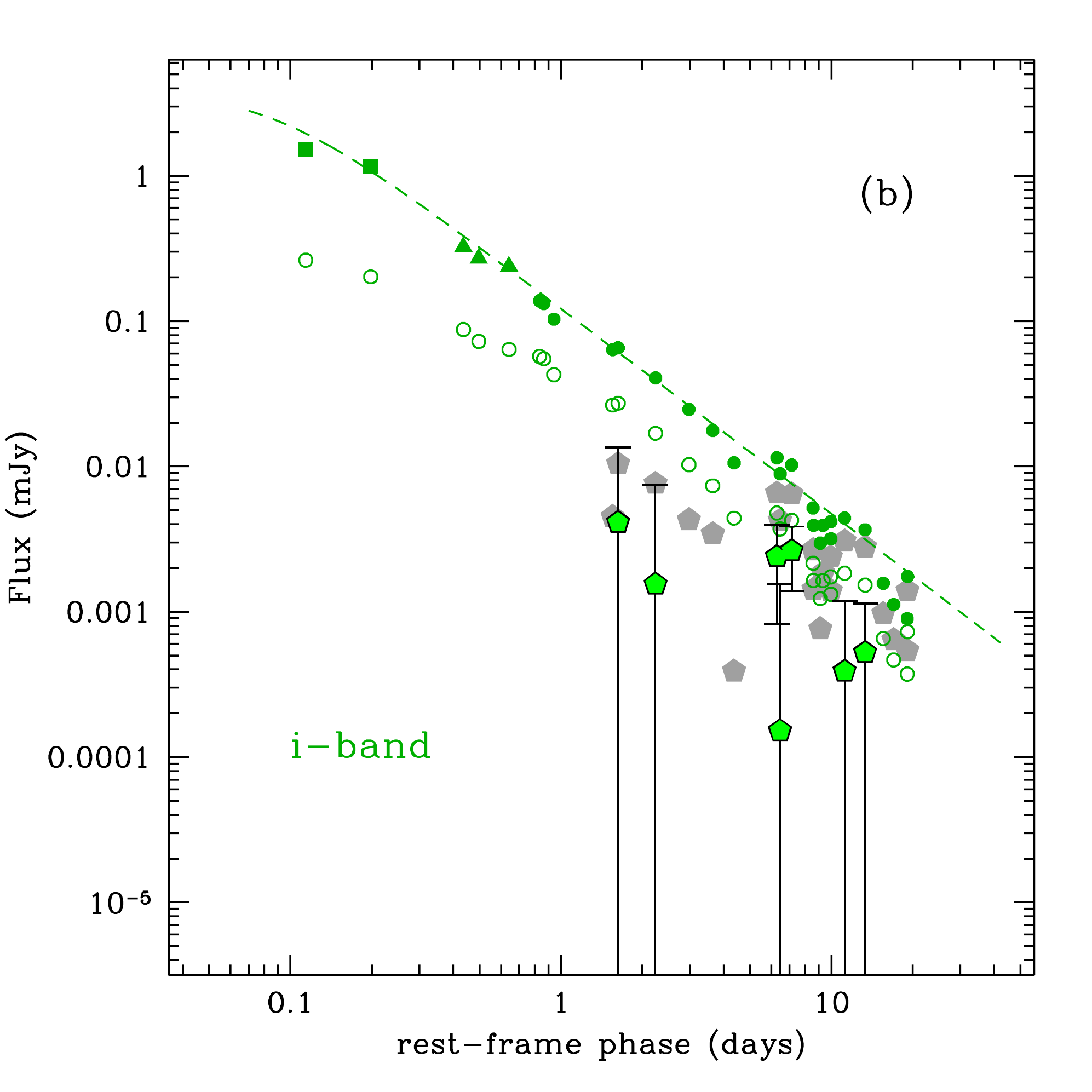}
     \includegraphics[width=8cm,height=8cm]{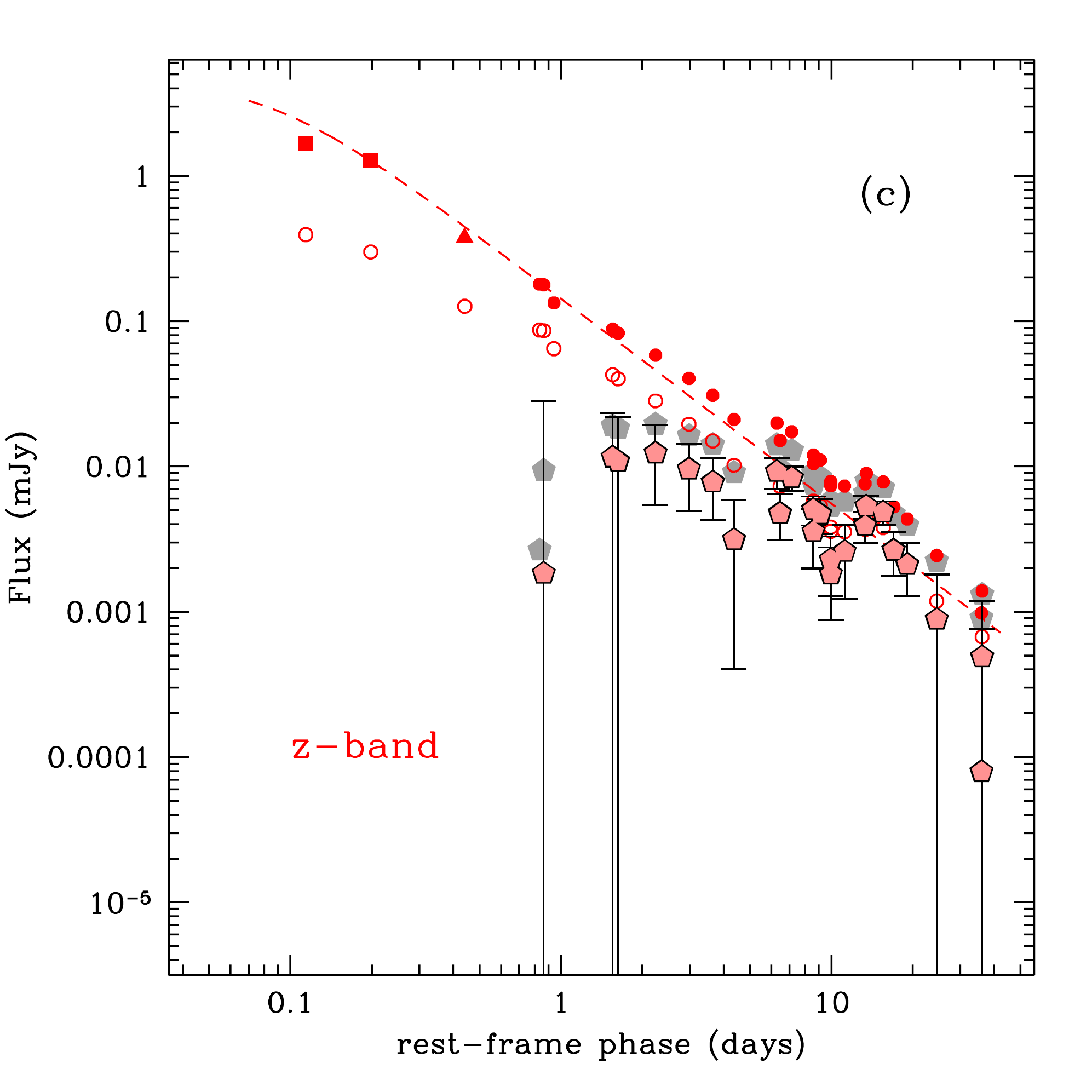}
     \includegraphics[width=8cm,height=8cm]{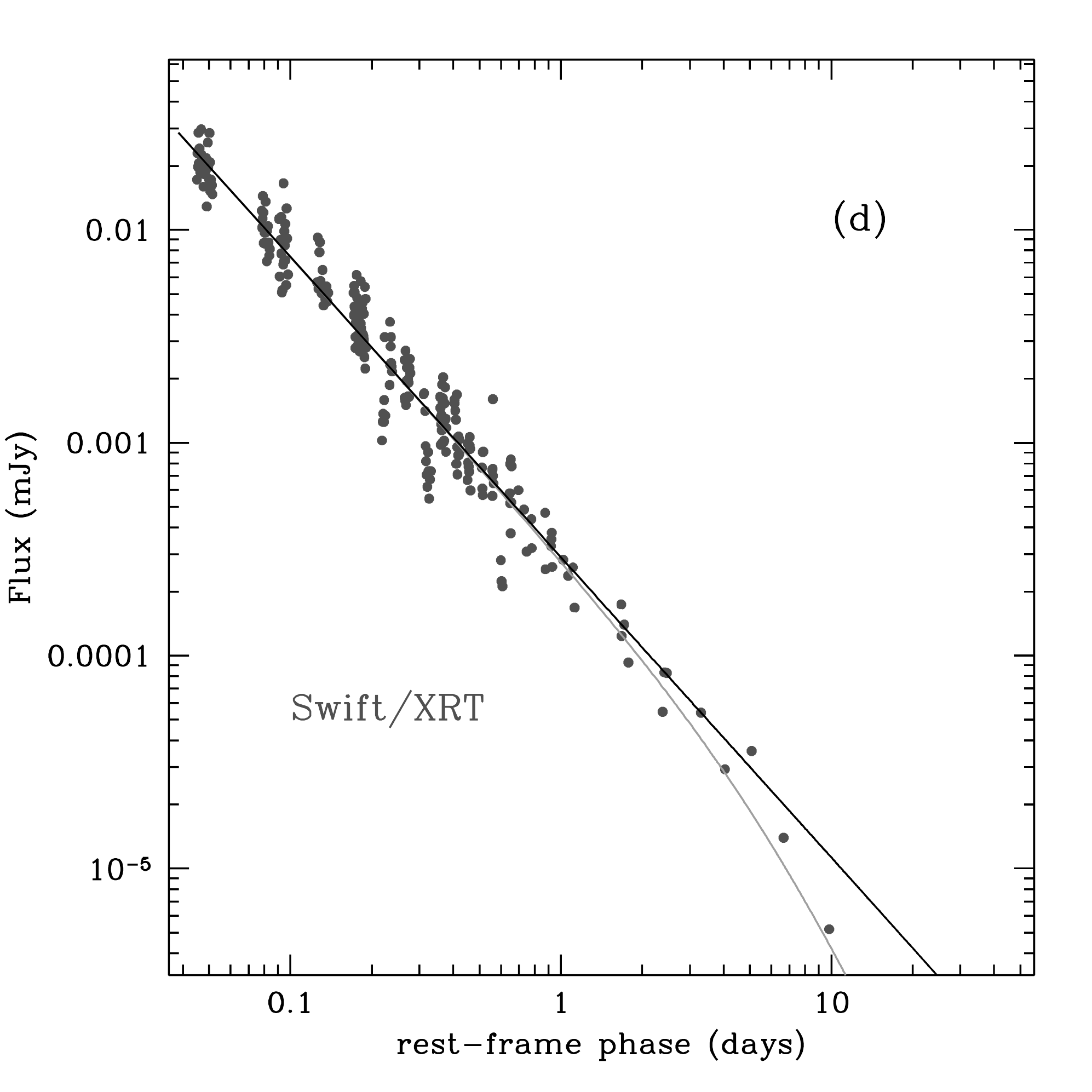}
   \caption{(a) Light curves of the GRB\,190114C counterpart in the optical $r$-band (open blue circles). All fluxes were corrected for host galaxy contribution, Galactic extinction and redshift (but not K-corrected). A correction for dust absorption along the GRB line of sight in its host galaxy was also applied  ($E(B-V) = 0.6$\,mag, phase $< 0.3$\,day, filled squares; $E(B-V) = 0.45$\,mag, 0.3 $<$ phase $<$ 0.7\,day, filled triangles;  $E(B-V) =  0.3$\,mag, phase $> 0.7$\,day, filled circles). The synchrotron model with no jet break ($t^{-1.4}$) is shown as a dashed blue curve at optical wavelengths and the corresponding SN component is represented by light-blue pentagons. If a jet break occurs at 7\,days, the SN component is shown as gray pentagons. (b) Same as (a) for the near-infrared $i$-band. Galaxy-subtracted points are in dark green and the SN points in light green. (c) Same as (a) for the near-infrared $z$-band. Galaxy-subtracted points are in red and the SN points in pink. (d) Light curves of the GRB\,190114C counterpart in X-rays (0.3--10\,keV {\it Swift}/XRT data, dark gray filled circles). The synchrotron model with no jet break ($t^{-1.4}$) is shown as a solid black curve. If a jet break occurs at 7\,days, the X-ray synchrotron model is shown as a solid gray curve. For clarity, uncertainties on data points were omitted.}
    \label{figLC2}%
    \end{figure*}

\subsection{Imaging}

We observed the field of GRB\,190114C  between  1.34 and $\sim 51.5$ days (in the observer frame) after the burst event with several facilities: the 3.58\,m New Technology Telescope (NTT), the 4.2\,m William Herschel Telescope (WHT), the 3.58\,m Galileo National Telescope (TNG), the 8.2\,m Very Large Telescope (VLT), and the Large Binocular Telescope (LBT, two twin 8.4\,m telescopes at Mt. Graham in Arizona, USA). 

We also observed the location of GRB\,190114C with the {\it Hubble Space Telescope (HST)} at five epochs 27--370\,days after the burst with the ACS/WFC and three optical/near-infrared (NIR) filters (F606W, F775W, and F850LP).  These frames  were reduced via {\tt astrodrizzle} to a final scale of $0.025''$\,pixel$^{-1}$.  Some data obtained at late times with the NIR filters F110W and F160W were presented by \cite{deUgarte}.  

All of our images were reduced following standard procedures, including debiasing and flat-fielding. The photometric calibration was carried out by using a set of field stars selected from the Pan-STARRS1 data archive\footnote{https://panstarrs.stsci.edu/}.

In Fig. \ref{figField} we show three images obtained with {\it HST}, VLT, and LBT at  similar epochs. GRB\,190114C is located close to the nuclear region of a galaxy (``A'') that interacts with a companion galaxy (``B'') located $\sim 1''$ to its northeast \citep{deUgarte}. The two galaxies are almost blended in the majority of our ground-based images, so their combined light must be subtracted in order to isolate the transient emission. Aperture photometry was used for all images, including those from {\it HST}, with a radius of $\sim 2.5''$ to include both galaxies. From the measured magnitudes (reported in Table~\ref{LogTab1}) we subtracted the combined flux of the two-galaxy system ($r = 21.66$\,mag, $i = 21.09$\,mag, and $z = 20.97$\,mag), as estimated from the latest {\it HST} observations. These galaxy-subtracted magnitudes are reported in Fig. \ref{figLC2}). No K-correction was applied, owing to the lack of simultaneous spectra covering the IR wavelength range. We note that the $r$-band light curve (Fig. \ref{figLC2}a) levels off after day $\sim$20, suggesting some extra background contribution ($\sim$0.001 mJy) that was not accounted for in our subtraction (this may have occurred for instance if a compact clump due to a star forming region is located right under the location of the GRB). This extra contribution was subtracted from the $r$-band points before computing the bolometric light curve.

\subsection{Spectroscopy}

We observed GRB\,190114C in the optical/NIR band with the LBT using the Multi-Object Double Spectrographs MODS-1 and MODS-2 (Pogge et al. 2010) in dual grating mode (grisms G400L and G670L) on 2019 Jan. 29  (mid-observation time of 13.25\,days after the burst trigger). For both MODS spectra we used the same instrumental configuration, consisting of a slit width of $1.2''$ and dual-grating mode, covering the spectral range 3200--9500\,\AA, and yielding spectral resolution $R = \lambda / \Delta \lambda \approx 1000$. Considering both MODS together, a total of 2.0\,hr of exposure time was obtained under moderately good seeing ($\sim 1.3$\arcsec), but at high airmass ($\sim 2$) given the low declination of the target.

Data reduction was carried out at the Italian LBT Spectroscopic Reduction Center\footnote{http://www.iasf-milano.inaf.it/software} by means of scripts optimized for LBT data adopting the standard procedure for long-slit spectroscopy with bias subtraction, flat-fielding, bad-pixel correction, sky subtraction, and cosmic-ray decontamination. Wavelength calibration (in air) was obtained using spectra of Hg, Ar, Xe, and Kr lamps, providing an accuracy of $\sim 0.08$\,\AA\ over the whole spectral range. Relative flux calibration was derived from the observations of a spectrophotometric standard star. 

We also obtained spectra of GRB\,190114C with the VLT/FORS2, using the low-resolution 300I-OG590 grism, at 19.17, 22.15, 27.19, and 51.2 observer days after the GRB detection. The complete journal of observations and the corresponding instrumental configurations are reported in Table \ref{tab:FORS} (see also Fig. A.1 in the appendix). The fourth-epoch spectrum was used as a template for the underlying host-galaxy system. In this spectrum, we see no flux excess at the wavelengths corresponding to the $r$-band filter (see Section 2.1). Therefore, we have not subtracted any extra flux from the spectra.  If any such extra background was present, its subtraction would make the spectrum redder still (see Section 3.2). We reduced the FORS2 data using \textsc{esorex} recipes that permitted us to first correct the raw science frames for the bias and the flat-field and then to determine the spectral extraction mask used for the wavelength calibration of the science and standard-star spectra. Thereafter, we obtained the two-dimensional wavelength- and flux-calibrated GRB spectra, and finally used the \textsc{IRAF}-\emph{apall} package for a more accurate background subtraction. The wavelength range covered in all of our observations is 6100--10,300\,\AA, which corresponds to the rest-frame wavelength range 4280--7227\,\AA.

We performed accurate flux calibration of our spectra using $RIz$ photometry obtained with FORS2, using a polynomial interpolating function to first fit the observed photometry, then convert to flux densities, and finally determine the correction term for each spectrum. We also corrected for the fact that the $R$ filter covers only $\sim 46$\% of the observed spectral emission.

\section{Results}

\subsection{X-ray and optical/NIR afterglow}

In Fig. \ref{figLC2} our optical and NIR observations are shown together with data in the same bands taken from the literature \citep{Magicnature2,misra21,jm20,kann19,im19,pd19,ki19,kumar19,ki19b,mazaeva19,kumar19b,watson19,bik19}, and with the {\it Swift}/XRT light curve in the observed range 0.3--10\,keV, as archived in the {\it Swift} repository\footnote{https://www.swift.ac.uk/xrt$\_$curves/00883832/}  \citep{evans09}.

Prior work on the very early afterglow phase of GRB\,190114C (i.e., during the first hours after explosion) found this to be dominated by a reverse-shock component (Laskar et al. 2019; Jordana-Mitjans et al. 2020). \citet{misra21} determined the presence of intrinsic absorption along the line of sight of $A_V = 1.9$--2.4\,mag (for a Milky Way extinction curve). They also analysed the radio and X-ray afterglow data from early times to $\sim 100$\,days, ignoring the optical data on account of the fact that they are a affected by host-galaxy and SN components. Under the assumption that a spectral cooling break is located between the radio and X-ray wavelengths, they found no clear consistency of the data with the standard fireball scenario unless  the shock microphysical parameters vary with time. The \citet{Magicnature2} analysis of the radio-to-X-ray afterglow in the first 3 days determined an intrinsic $A_V = 1.8$\,mag, consistent with that found by \citet{misra21}.  In this paper, we focus on the SN component and aim at characterising it in the context of other nearer GRB-SNe and stripped-envelope core-collapse SNe. Our decomposition of the optical galaxy-subtracted data into afterglow and SN components takes into account the previous analyses but introduces the element of intrinsic absorption variation.

In order to decompose the SN\,2019jrj light from the galaxy-subtracted optical/NIR photometry, we constructed a model for the multiwavelength afterglow. Following previous authors \citep{Magicnature2,fraija19a,misra21}, we assumed that a forward shock propagating in an external medium of uniform density is responsible for the X-ray and optical/NIR emission starting a few hours after explosion. We adopted the classical fireball formalism \citep{zm04,kz15}; in particular, the electron energy power law has the form $dE/d\gamma \propto \gamma^{-p}$; $\nu_m$  and $\nu_c$ indicate the characteristic synchrotron frequencies related to the minimum electron energy and to the cooling energy, respectively. The synchrotron flux depends on time and frequency as $f(t) \propto \nu^{-\beta} t^{-\alpha}$. 

The MAGIC Collaboration analysis of the multiwavelength afterglow at early epochs ($<3$ observer days;  \citealt{Magicnature2}, see their Extended Data Fig. 6) shows that the frequency $\nu_m$ crosses the optical band $<2$ rest-frame hours after the explosion. On the other hand, the cooling frequency $\nu_c$ is still above the X-ray frequencies at $\sim 2$ rest-frame days. Since $\nu_c$ scales as $t^{-1/2}$, it does not cross the X-ray band in the time interval covered by our optical monitoring. Therefore, from the X-ray spectral index $\beta_X = 0.94$ reported in the {\it Swift}/XRT repository\footnote{https://www.swift.ac.uk/xrt$\_$spectra/00883832/} \citep{evans09}, we derive an electron energy power-law index $p = 2\beta_X + 1 = 2.88$. The decay of the X-ray light curve is then given by $\alpha = (3/4) (p - 1) = 1.41$, consistent with the value of $\alpha_X = 1.344 \pm 0.003$ fitted by \citet{misra21}. This is shown by a solid line in Fig. \ref{figLC2}d.

Since there is no cooling break between the optical and X-ray bands, the optical spectral slope and time decay past the $\nu_m$ passage must be the same as in X-rays. However, the thin synchrotron spectrum that fits the X-ray emission over-predicts the optical galaxy-subtracted flux (dashed curves in Fig. 2a,b,c). This was interpreted by the MAGIC Collaboration as evidence of dust extinction along the line of sight to the GRB on top of the Galactic one. Their estimate for this intrinsic extinction is $A_V = 1.83 \pm 0.15$\,mag, broadly in agreement with other independent analyses of the early-time afterglow spectral energy distribution \citep{laskar19,misra21}. However, this value exceeds the amount of extinction that is necessary to correct our data after 0.7 rest-frame days (indeed, their optical flux distributions at $\sim 0.7$ and $\sim 2$ rest-frame days also seem overcorrected). For a reddening of $E(B-V) \approx 0.3$\,mag, assuming it takes place entirely at the redshift of the GRB (likely dust extinction in the host galaxy), our optical data after 0.7\,day match the synchrotron prediction (see Fig. \ref{figLC2}a,b,c). Prior to that, the optical data are still below the synchrotron time curve and present a flattening toward the earlier epochs that cannot be accounted for only by the passage of $\nu_m$. A higher intrinsic reddening, gradually decreasing from $E(B-V) = 0.6$\,mag to $E(B-V) = 0.3$\,mag, is necessary to correct the data between 0.1 and 0.7 rest-frame days, consistent with the MAGIC Collaboration finding. This suggests that the intrinsic dust extinction has possibly varied with time. Note that the Galactic extinction curve used by us \citep{cardelli89}, and the Large Magellanic Cloud (LMC) extinction curve used by MAGIC Collaboration to evaluate the intrinsic extinction, coincide at the rest-frame wavelengths of our observations \citep{pei92}.

As also noted by \citet{misra21}, the X-ray light curve  may decay at a quicker rate after 1 rest-frame week, probably owing to the presence of a jet break, which should be achromatic. Because of the likely emergence of an SN component in the same time interval, it is difficult to establish directly whether the optical light curve presents a similar jet break. However, if the optical synchrotron light curve steepened simultaneously with the X-rays, the residuals of its subtraction from the galaxy-subtracted points, representing the SN component, would depend critically on this break time. We have thus modeled the steepening light curve assuming a post-break decay index of $\alpha_2 = p = 2.88$ and a range of time breaks.  The X-ray light curve is consistent with any time break larger than $\sim 7$ rest-frame days. In  Fig. \ref{figLC2}d we show the X-ray light-curve fit for the two extreme cases of a break at $\sim 7$\,days and at $\sim 100$ rest-frame days, the latter virtually coinciding with no break during the X-ray monitoring.

%

%--------------------------------------------------------------------

\subsection{Supernova component}

\begin{figure}[!ht]
   \centering
    \includegraphics[width=9.0cm,height=8.0cm]{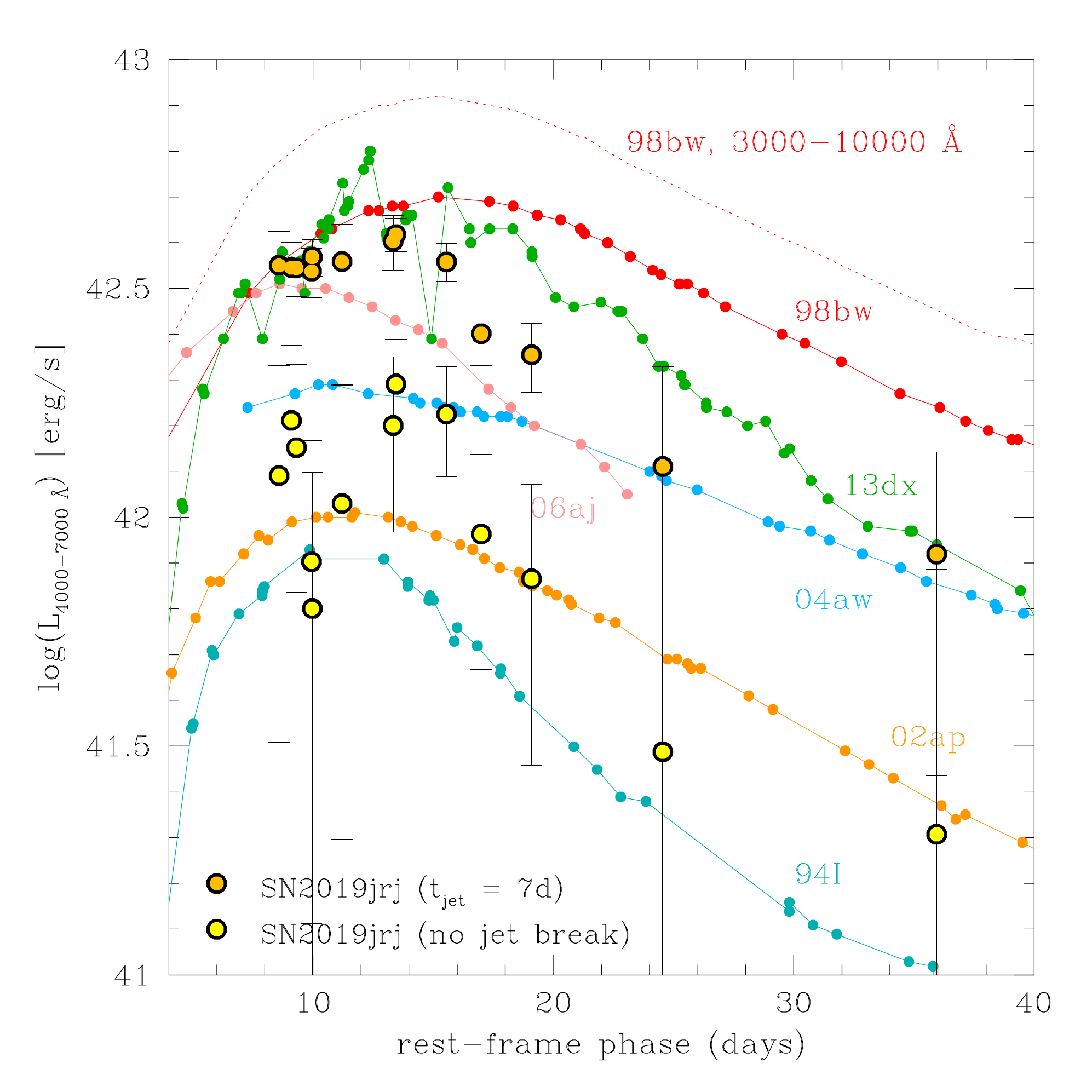}
   \caption{Light curves of SN\,2019jrj computed in the 4000--7000\,\AA\  range under the two assumptions that a jet break occurs at 7 rest-frame days or does not occur until at least several months after the GRB. The light curves of other core-collapse SNe computed in the same wavelength range from available photometry \citep{galama98,mks99,patat01,delia15,toy16,mazzali21,pian06,ferrero06,tau06,foley03,tomita06} are shown, as well as the light curve of SN\,1998bw in the 3000--10,000\,\AA\ range (dotted red curve). For clarity, uncertainties are shown only for SN\,2019jrj.}
    \label{figBol}%
    \end{figure}

Although the decomposed SN\,2019jrj curves cover a limited wavelength range ($\sim 4000$--7000\,\AA\ in the rest frame), we have attempted to construct a pseudobolometric light curve in this interval by integrating the SN\,2019jrj signal in  the $riz$ bands and adding flux blueward and redward of this range by extrapolating the spectral flux to 4000\,\AA\ and 7000\,\AA, respectively, using a constant.  In Fig. \ref{figBol} we report two pseudobolometric curves computed in this way, under the two extreme assumptions that a jet break occurs at 7 rest-frame days, or does not occur until at least several months after the GRB. These two curves provide the most probable range of the pseudobolometric luminosity of SN\,2019jrj. For comparison, we show the bolometric light curves of other core-collapse SNe computed in the same wavelength interval. A break at 7 rest-frame days causes the SN component to have a peak luminosity similar to that of SN\,2013dx, while absence of a jet break (i.e., jet break occurring at times later than a few months) causes the SN component to have a peak luminosity comparable to less luminous core-collapse SNe, in particular SN\,2004aw (Fig. \ref{figBol}).

\begin{figure*}[!ht]
 %  \centering
 \hspace{-0.4cm}
    \includegraphics[width=6.25cm]{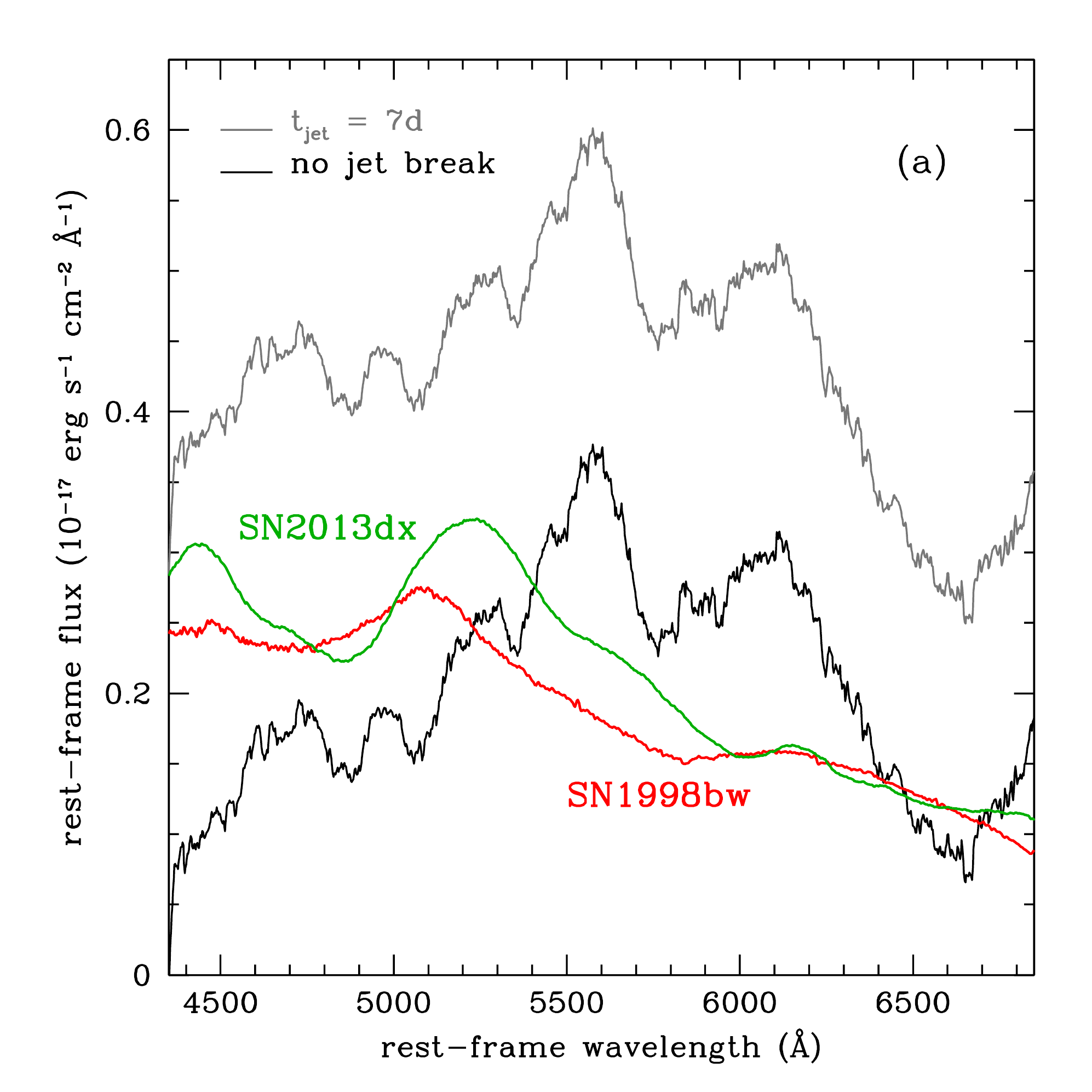}
    \includegraphics[width=6.25cm]{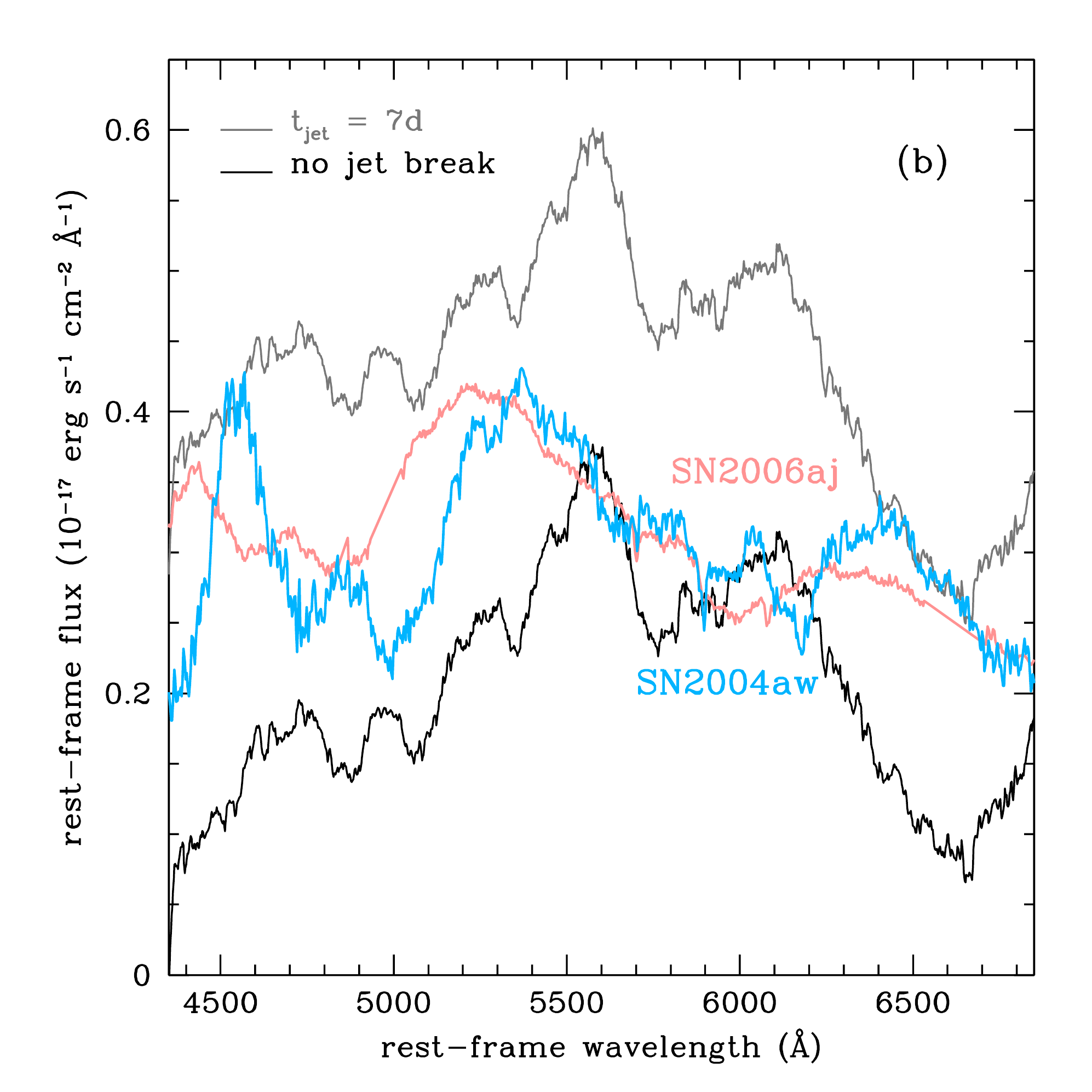}
    \includegraphics[width=6.25cm]{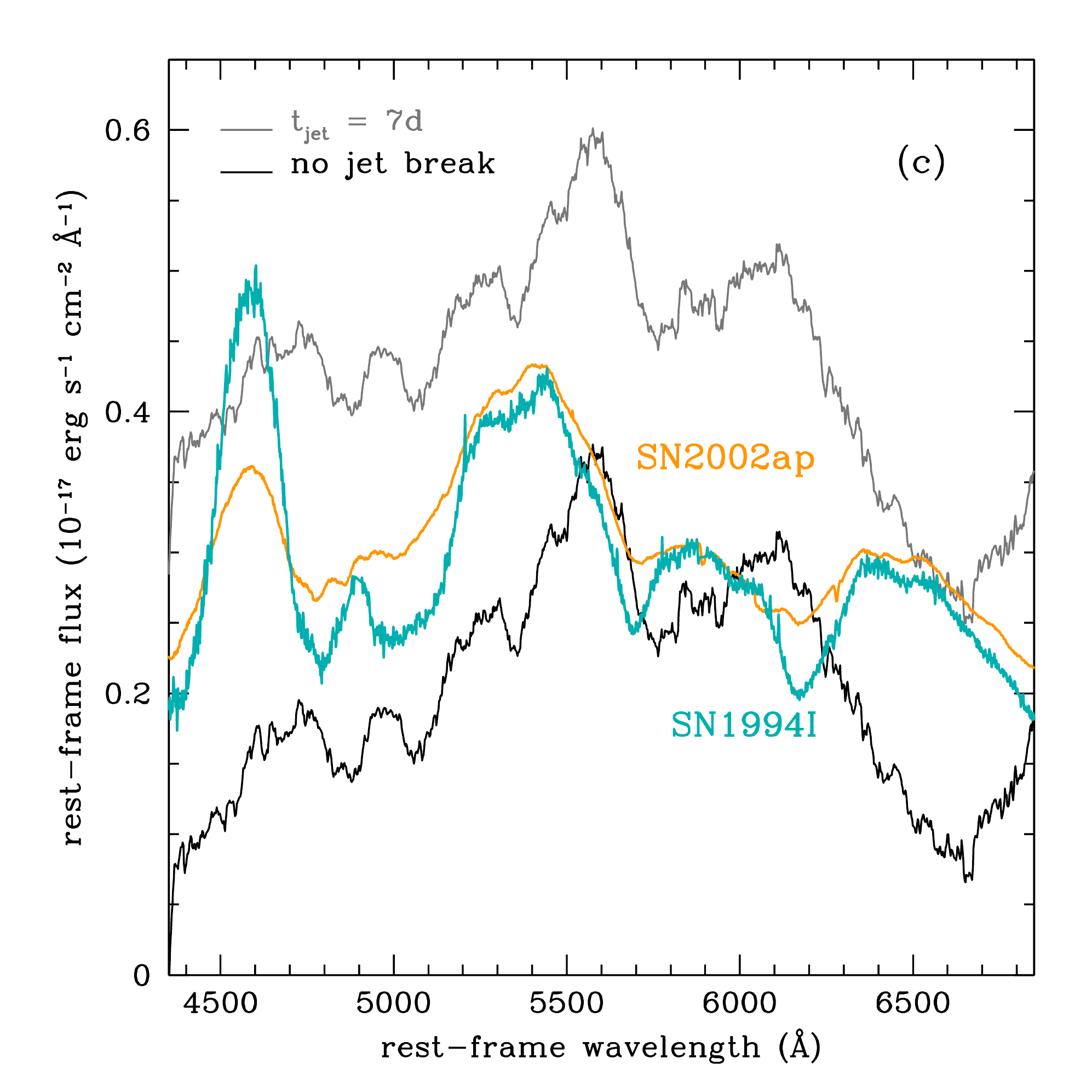}
  \caption{(a) Spectra of SN\,2019jrj at maximum light obtained under the assumption of a jet break at 7 rest-frame days (gray upper curve) and no break (black lower curve), and corrected for Galactic \citep[$E(B-V)_{\rm Gal} \approx 0.013$\,mag;][]{sf11}) and line-of-sight reddening at the source ($E(B-V) = 0.3$\,mag). For comparison, maximum-light spectra of GRB-SNe 1998bw \citep{patat01} and 2013dx \citep{delia15,mazzali21} are shown. These are dereddened for Galactic extinction, reduced to the rest frame, and scaled in flux to match the SN\,2019jrj flux level. (b) Same as in panel (a), compared with maximum-light spectra of the core-collapse Type Ic SNe 2004aw \citep{tau06} and 2006aj (associated with an XRF; Pian et al. 2006; Mazzali et al. 2006b). (c) Same as in panel (a), compared with maximum-light spectra of the core-collapse Type Ic SNe 1994I \citep{filippenko95,foley03} and 2002ap \citep{mazzali02,foley03}.}
    \label{figSpec_XX}%
  \end{figure*}

Among our spectra (see Table \ref{tab:FORS}), we concentrate our analysis on that taken 13.45 rest-frame days after the explosion (approximately corresponding to the maximum brightness of SN\,2019jrj), which has the highest signal-to-noise ratio. Even so, individual features cannot be unambiguously detected and atomic species cannot be identified.  Therefore, our considerations below are based only on  the general spectral shape and appearance. After correcting this spectrum for Galactic \citep[$E(B-V)_{\rm Gal} \approx 0.013$\,mag;][]{sf11} and intrinsic (with $E(B-V) = 0.3$\,mag) extinction in the rest frame and subtracting the afterglow component, the residuals are inconsistent with the spectrum at maximum light of the most luminous GRB-SNe 1998bw and 2013dx (panel (a) in Fig. \ref{figSpec_XX}).   

Comparison with maximum-light spectra of less-luminous core-collapse SNe is more satisfactory (panels (b) and (c) of Fig. \ref{figSpec_XX}). Since both SN luminosity and spectral shape depend on temperature, we conclude that SN\,1998bw must be regarded only as an extreme analogue of SN\,2019jrj, while its best proxies are in fact less luminous core-collapse SNe with narrower lines (i.e., lower photospheric velocities) and without a detected GRB (SN\,2002ap, SN\,2004aw). This in turn suggests that, while a jet break that occurred as early as 7\,days after GRB explosion is formally acceptable, it probably took place later. Along the same line of reasoning, if the break observed in the X-ray light curve at 7\,days is wavelength-dependent, it is however unlikely that it occurs much earlier in the optical, as in this case the resulting SN would be significantly more luminous than SN\,1998bw, which would only worsen the incompatibility with the spectral shape of SN\,2019jrj at maximum brightness.

Note that, if the optical light were absorbed by a constant intrinsic absorption $A_V \approx 2$\,mag  \citep{Magicnature2,misra21}, the SN peak luminosity under the assumption of a jet break at 7\,d would be around $10^{43}$\,erg\,s$^{-1}$. Such a luminosity is more typical for superluminous supernovae for which a long-lived blue spectrum is commonly observed \citep[e.g.][]{galyam19}. This is inconsistent with the relatively red spectral shape observed for SN\,2019jrj.

The best analogue of SN\,2019jrj in luminosity level, light-curve shape, and spectral line width appears to be SN\,2004aw, suggesting that these two SNe may have similar physical properties ---  a  synthesized $^{56}$Ni mass of $M_{\rm Ni} \approx 0.2$\,M$_{\odot}$, an ejecta mass of $M_{\rm ej} \approx 3$--5\,M$_{\odot}$, a kinetic energy of $E_{\rm kin} \approx (3-6) \times 10^{51}$\,erg, and a progenitor mass of $\sim 25$\,M$_{\odot}$ \citep{mazzali17}.  We stress, however, that these quantities should be regarded as rather uncertain, considering the relatively low quality of the light curve and spectra of SN\,2019jrj and our poor ability to constrain the time of the jet break.

We have also performed an analysis of the pseudobolometric light curve using the analytical model developed by \citet{Arnett1982} for Type Ia SNe. This model can be applied to core-collapse Type Ic SNe with some caveats: first, the model assumes spherical symmetry for the SN ejecta and a constant opacity $\kappa$ (fixed to 0.07\,cm$^2$\,g$^{-1}$) throughout the entire ejecta; second, it assumes that the total amount of nickel is concentrated at the center of the ejecta, an assumption that for highly-rotating progenitor stars is not entirely true \citep{izzo19,ashall19}. The model provides an estimate of the total kinetic energy of the SN ejecta, given the expansion velocity measured from P~Cygni absorption of spectral features around the peak brightness of the SN\footnote{Measuring the total kinetic energy using the spectra at maximum light is highly uncertain. As shown by \citet{mazzali17}, a change in kinetic energy of 50\% is only visible in the spectra one week after the explosion, and not at maximum light. Therefore, we take 50\% of the estimated value as the lower limit of the uncertainty in our measurement.}, as well as an estimate of the total amount of $^{56}$Ni synthesized in the explosion. The application of the Arnett formalism to our pseudobolometric light curves, assuming an average photospheric velocity of 15,000\,km\,s$^{-1}$, yields consistent results with those obtained above by following the analogy with SN\,2004aw.

\begin{table*}[!ht]
      \caption[]{Observed magnitudes (AB system) of the transient prior to subtraction of the two-galaxy system, and not corrected for Galactic extinction. Time since burst ($\Delta t$) is in the observer frame.}
         \label{LogTab1}
         \centering
\begin{minipage}{17cm}
\begin{tabular}{ccc|ccc|ccc}
\hline \hline
 $\Delta$t & Magnitude & Telescope & $\Delta t$ & Magnitude & Telescope & $\Delta t$ & Magnitude & Telescope\\
 (d) & (err) & & (d) & (err) & & (d) & (err)&\\
 \hline \hline
\multicolumn{3}{c|}{$g$}   & 227.13 & 21.66 (0.07) & HST & \multicolumn{3}{c}{$z$} \\
12.97 & 22.33 (0.08) & TNG & 370.57 & 21.66 (0.07) & HST & 1.34 & 18.84 (0.07) & NTT \\  
13.27 & 22.37 (0.07) & LBT & \multicolumn{3}{c|}{$i$}    & 8.96 & 20.27 (0.10) & WHT \\
\multicolumn{3}{c|}{$r$}   &  1.34 & 19.24 (0.05) & NTT  & 10.17 & 20.34 (0.06) & VLT \\
1.34 & 19.94 (0.05) & NTT  & 8.96 & 20.65 (0.06) & WHT	 & 12.23 & 20.50 (0.04) & VLT \\
8.96 & 21.33 (0.06) & WHT  & 10.17 & 20.69 (0.04) & VLT	 & 12.23 & 20.55 (0.10) & NTT \\
10.17 & 21.35 (0.04) & VLT & 12.22 & 20.87 (0.05) & VLT  & 12.97 & 20.53 (0.07) & TNG \\ 
12.22 & 21.40 (0.05) & VLT & 12.23 & 20.92 (0.09) & NTT  & 14.18 & 20.64 (0.05) & VLT \\ 
12.23 & 21.40 (0.05) & NTT & 12.97 & 20.96 (0.04) & TNG  & 15.97 & 20.66 (0.11) & TNG \\ 
12.97 & 21.42 (0.05) & TNG & 13.27 & 20.92 (0.05) & LBT  & 19.00 & 20.65 (0.07) & TNG \\ 
13.27 & 21.45 (0.05) & LBT & 14.17 & 20.91 (0.05) & VLT  & 19.20 & 20.60 (0.07) & VLT \\ 
14.17 & 21.43 (0.05) & VLT & 15.97 & 20.90 (0.05) & TNG  & 22.15 & 20.64 (0.07) & VLT \\ 
15.97 & 21.45 (0.13) & TNG & 19.00 & 20.93 (0.04) & TNG  & 24.22 & 20.74 (0.08) & NTT \\ 
19.00 & 21.45 (0.10) & TNG & 22.15 & 21.02 (0.05) & VLT  & 27.19 & 20.78 (0.08) & VLT \\ 
19.17 & 21.47 (0.03) & VLT & 24.22 & 21.04 (0.04) & NTT  & 34.98 & 20.86 (0.10) & TNG \\ 
22.15 & 21.44 (0.03) & VLT & 27.19 & 21.05 (0.07) & VLT  & 51.20 & 20.95 (0.08) & VLT \\ 
24.22 & 21.46 (0.05) & NTT & 34.98 & 21.13 (0.12) & TNG  & 51.44 & 20.94 (0.08) & VLT \\ 
27.19 & 21.46 (0.03) & VLT & 51.20 & 21.20 (0.12) & VLT  & 325.37 & 20.97 (0.09) & NTT \\
34.98 & 21.48 (0.15) & TNG & 325.36 & 21.17 (0.04) & NTT & \multicolumn{3}{c}{$F850LP$}\\
51.20 & 21.48 (0.10) & VLT & \multicolumn{3}{c|}{$F775W$}& 27.27 & 20.82 (0.09) & HST\\
325.34 & 21.51 (0.09) & NTT & 27.27 & 20.91 (0.08) & HST & 56.37 & 20.90 (0.07) & HST \\
\multicolumn{3}{c|}{$F606W$}& 165.03 & 21.05 (0.07) & HST & 227.13 & 20.97 (0.08) & HST \\
27.27 & 21.54 (0.07) & HST	& 227.13 & 21.09 (0.08) & HST & 370.57 & 20.97 (0.08) & HST \\
165.03 & 21.63 (0.06) & HST	& 370.57 & 21.09 (0.09) & HST & & & \\

\hline
\end{tabular}
\end{minipage}
\end{table*}

\begin{table}[!ht]
 \caption{Summary of spectroscopic observations.}
 \label{tab:FORS}
 \centering
 \begin{tabular}{lccccc}
  \hline
  Day & Phase$^{a}$ & Slit  & T$_{\rm exp}$ & Seeing & Tel./Instr. \\
  (2019) & (Day) & (\arcsec) & (s) & (\arcsec) & \\
  \hline
 Jan 28 & 9.30 & 1.0 & 8$\times$900 & 1.30 & LBT/MODS\\
 Feb 3 & 13.45 & 0.7 & 3$\times$900 & 1.26 & VLT/FORS2\\
 Feb 6 & 15.55 & 1.0 & 3$\times$900 & 0.70 & VLT/FORS2\\
 Feb 11 & 19.09 & 1.0 & 3$\times$900 & 0.90 & VLT/FORS2\\
 May 6 & 35.94 & 1.00 & 3$\times$900 & 0.65 & VLT/FORS2\\
  \hline
 \end{tabular}
 \scriptsize{{\it a})~~Phases with respect to the time of burst ($T_{\rm 0}$) are in the rest frame.~~~~~~~~~~~~~~~~~~~~~~~~~~~~}
\end{table}

\section{Discussion}

Our  analysis of GRB\,190114C/SN\,2019jrj, focusing both on the multiwavelength afterglow and on the SN emission, led to the following main conclusions.

GRB\,190114C was very energetic ($E_{\rm iso} \approx 3 \times 10^{53}$\,erg;  \citealt{Magicnature2}), and it was one of the most energetic GRBs with an associated SN. With a minimum jet break time of 7\,days, estimated from analysis of the X-ray light curve, GRB\,190114C had an opening angle of at least $13^\circ$. This corresponds, for a uniform medium density of $\sim 1$\,cm$^{-3}$, to a corrected energy output of $\sim 2 \times 10^{51}$\,erg. If the jet break occurred later, this translates into a larger energy. 

Accordingly, the luminosity of SN\,2019jrj may be as large as that of luminous GRB-SNe (particularly SN\,2013dx) or as low as that of stripped-envelope core-collapse SNe 2004aw and 2002ap (not accompanied by a detected high-energy event; Taubenberger et al. 2006; Tomita et al. 2006). This underlines the need for late-time accurate observations of low-redshift GRBs at all wavelengths,  to establish the presence of a possible jet break, which is in turn crucial to determine accurately the GRB intrinsic energy and the luminosity of the accompanying SN.

Interestingly, SN\,2019jrj does not resemble spectroscopically the most energetic (i.e., with the broadest absorption lines) GRB-SNe, like the prototypical SN\,1998bw. Its maximum-light spectrum is rather more similar to that of core-collapse SNe having less broad lines and no detected accompanying GRB, like SNe\,2004aw and 2002ap. This suggests that less luminous and less energetic SNe may also be viable GRB progenitors, and may still produce high-energy events of substantial energy.

Decreasing absorption by dust at the source redshift at epochs earlier than 1 observer day, as derived from our combined analysis of the optical and X-ray  afterglow light curves, is consistent with the drop of intrinsic neutral hydrogen absorption by a factor of 2 around the same time, as deduced from analysis of the X-ray spectra \citep{campana21}. The coherent behavior derived from these independent analyses of different datasets reinforces the case for a time-variable absorber composed by both dust and gas, although its origin is not known. The seeming contradiction of assuming a uniform medium density in our model in the presence of variable intrinsic absorption is mitigated by the fact that this variation is mild  and by our ignorance of the absorber nature and geometry.  

It has been suggested that the afterglow of GRB\,190114C could originate not only from synchrotron radiation but also from hadronic processes involving photomeson interaction, that may play an important role in the formation of the gamma-ray spectrum up to TeV energies \citep{Magicnature1,Magicnature2,derishev19,fraija19b,slf20}. If this is the case, the dissipation mechanism responsible for the acceleration of electrons up to high energy may also be responsible for the acceleration of protons to high energy and produce detectable high-energy neutrinos. In fact, the high-energy protons interact with photons, producing charged and neutral pions \citep{guetta15}. The pion decay products include leptons and photons:
\begin{align}
{\pi}^+ &\to \mu^+ + \nu_{\mu} \to e^+ +\nu_e+ \bar{\nu}_{\mu} + \nu_{\mu}\, , \label{eq:36}\\
{\pi}^- &\to \mu^- + \bar{\nu}_{\mu} \to e^- +\bar{\nu}_e+\nu_{\mu} + \bar{\nu}_{\mu}\, , \label{eq:37}\\
{\pi}^0 &\to \gamma + \gamma \, . \label{eq:38}
\end{align}

The TeV fluence of GRB\,190114C has been estimated to be $f_{\rm TeV} \approx 0.1f_{\gamma} \approx 4.43 \times 10^{-5}$\,erg\,cm$^{-2}$ and the TeV power-law spectrum index to be $\sim -2$ \citep{Magicnature2}. The particle spectra at the source may be obtained through a Monte Carlo simulation, as shown in Fig. 1 of \citet{fasano21}, where the energy spectrum of the interacting protons, as well as the secondary particles emerging from the neutral pion decay of the interaction, are presented. The ordinate axis is in arbitrary units, as these spectra are not normalized. In order to determine a normalization factor we can assume that all of the TeV emission of GRB\,190114C is due to a hadronic mechanism \citep{yacobi14}. In this case (Gagliardini et al. 2021, in prep.), the expected neutrino flux from this source is $f_{\nu} \approx 0.5f_{\rm TeV} \approx 2 \times 10^{-5}$\,erg\,cm$^{-2}$. Considering the effective area of IceCube corresponding to the declination of the source ($\delta \approx -26^{\circ}$), the expected average number of upward muon detections for GRB\,190114C is 0.04. Therefore, the nondetection of neutrinos from IceCube is not surprising. However, sources that have a declination similar to that of GRB\,190114C are very good targets for the KM3Net neutrino telescope (because its effective area is maximal for sources with negative declination) that will be built in the Mediterranean Sea.

\begin{acknowledgements}
  We thank the anonymous referee for valuable comments and suggestions that improved the paper. A.M., M.G.B., P.D.A., S.C., and G.T. acknowledge support from ASI grant I/004/11/5. P.D.A., M.D.V., E.Pa., S.Sa., and S.P. acknowledge support from PRIN-MIUR 2017 (grant 20179ZF5KS). L.I. was supported by the VILLUM FONDEN (project numbers 16599 and 25501). A.V.F.'s research is supported by the Christopher R. Redlich Fund, the Miller Institute for Basic Research in Science (in which he is a Senior Miller Fellow), and many individual donors. D.A.K. acknowledges support from Spanish National Research Project RTI2018-098104-J-I00 (GRBPhot). D.W. is supported by Independent Research Fund Denmark grant DFF–7014-00017. The Cosmic Dawn Center is supported by the Danish National Research Foundation under grant 140. A.R. acknowledges support from the project Supporto Arizona \& Italia. The LBT is an international collaboration among institutions in the United States, Italy, and Germany. LBT Corporation partners are The University of Arizona on behalf of the Arizona Board of Regents; Istituto Nazionale di Astrofisica, Italy; LBT Beteiligungsgesellschaft, Germany, representing the Max-Planck Society, The Leibniz Institute for Astrophysics Potsdam, and Heidelberg University; The Ohio State University, representing OSU, University of Notre Dame, University of Minnesota, and University of Virginia. This work made use of data supplied by the UK Swift Science Data Centre at the University of Leicester.
\end{acknowledgements}


\appendix

\section{Spectral sequence}

  \begin{figure}[!ht]
   %\centering
   \vspace{-2.0cm}
   \hspace{-1.5cm}
    \includegraphics[width=20.5cm,height=16cm]{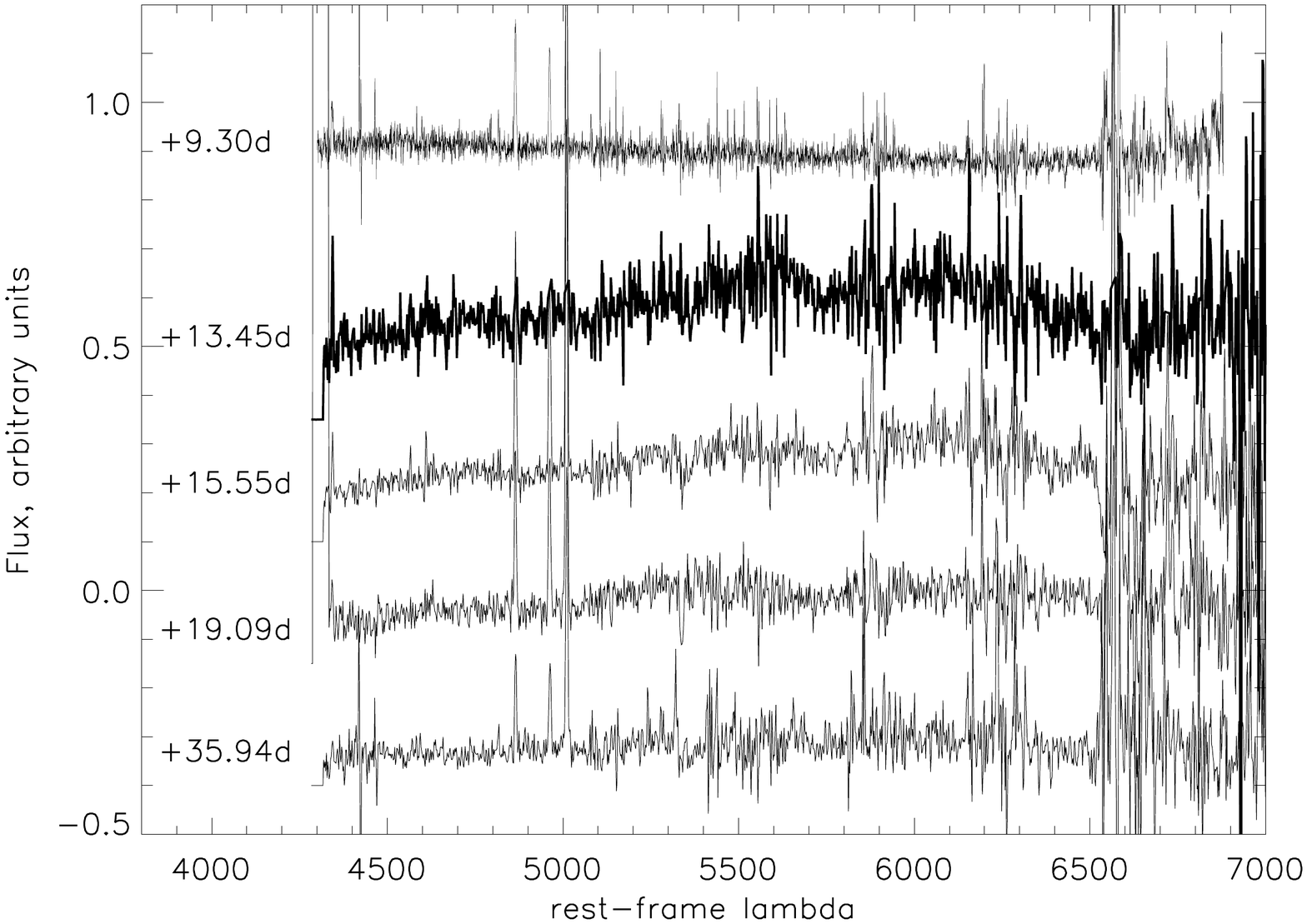}
   \caption{SN\,2019jrj spectral sequence obtained with LBT and VLT telescopes between 9.3 and $\sim$ 36 rest-frame days (see Table 2 and Section 2.2 in the main text for more details).}
    \label{spectsequence}%
    \end{figure}

\end{document}